\documentclass[twocolumn,aps,showpacs,superscriptaddress]{revtex4}
\usepackage[latin9]{inputenc}
\usepackage{amsmath}
\usepackage{amssymb}
\usepackage{graphicx}
\usepackage{esint}
\usepackage{amsfonts}
\usepackage{mathrsfs}
\usepackage{bm}
\usepackage{bbold}
\usepackage[usenames]{color}
\usepackage{array}

\newcommand{\be}{\begin{eqnarray}}
\newcommand{\ee}{\end{eqnarray}}
\newcommand{\beq}{\begin{equation}}
\newcommand{\eeq}{\end{equation}}
\newcommand{\bal}{\begin{align}}
\newcommand{\eal}{\end{align}}

\newcommand{\br}{{\bf r}}

\newcommand{\bk}{{\bf k}}

\newcommand{\bbone}{{\mathbb{1}}}
\DeclareMathAlphabet{\mathcalligra}{T1}{calligra}{m}{n}
\DeclareMathAlphabet{\mathpzc}{OT1}{pzc}{m}{it} 
\begin{document}
\title{
Hund
interaction, spin-orbit coupling and the mechanism of superconductivity in
strongly
hole-doped iron pnictides}
\author{Oskar Vafek}
\affiliation{Department of Physics and National High Magnetic Field Laboratory,
Florida State University, Tallahassee, Florida 32306 USA}

\author{Andrey V.~Chubukov}
\affiliation{School of Physics and Astronomy, University of Minnesota, Minneapolis,
MN 55455, USA}

\begin{abstract}
We present a novel mechanism of $s-$wave pairing in Fe-based superconductors.
The mechanism involves holes near $d_{xz}/d_{yz}$ pockets {\it only}
and is applicable primarily to strongly hole doped materials.
We argue that as long as the renormalized Hund's coupling $J$ exceeds the renormalized inter-orbital Hubbard repulsion $U'$, any finite spin-orbit coupling
gives rise to s-wave superconductivity. This holds even at weak coupling and regardless of the strength of the intra-orbital Hubbard repulsion $U$.
The transition temperature grows as the hole density decreases. The pairing gaps are four-fold symmetric, but anisotropic, with the possibility of eight accidental nodes along the larger pocket. The resulting state
is consistent with the experiments on KFe$_2$As$_2$.
\end{abstract}

\date{\today}
\maketitle

{\it {\bf  Introduction}}.~~~ The pairing mechanism in iron-based superconductors (FeSCs) remains
 the subject of intense debates~\cite{review}.
  A common scenario is that superconductivity
  (SC) is mediated by anti-ferromagnetic spin fluctuations, which are enhanced by the presence Fermi pockets of both hole and electron type~\cite{review,mazin,scal}.
   This scenario yields an
   $s-$wave pairing amplitude
    with opposite sign
     on hole and electron pockets.
  Such an $s^{+-}$ gap structure is consistent with experiments on moderately doped FeSCs which contain hole and electron pockets.

However, SC is also observed in strongly doped FeSCs with only hole or only electron pockets~\cite{KFeAs,ARPES_KFEAS,shin,louis,louis_2,raman,sh,KFeSe,neutr_kfese}.
For these systems, it is not clear why spin fluctuations should be strong enough to overcome Coulomb repulsion.

In this paper we focus on the systems with only hole pockets,
such as K$_x$Ba$_{1-x}$Fe$_2$As$_2$.
For
 KFe$_2$As$_2$, angle-resolve photoemission (ARPES) experiments show that only hole pockets are present~\cite{ARPES_KFEAS,shin}. Yet, $T_c\approx 3K$ in KFe$_2$As$_2$ and
increases as $x$ decreases.
The electronic structure of  KFe$_2$As$_2$ consists of three hole pockets centered at $\Gamma$ and hole ``barrels'' near $M = (\pi,\pi)$
in the Brillouin zone corresponding to a single Fe-As layer
with two Fe atoms per primitive unit cell.
The inner and the middle pockets at $\Gamma$ have the symmetry of
$d_{xz}$ and $d_{yz}$ orbitals, while the outer has the symmetry of the $d_{xy}$ orbital~\cite{scal}.

There is no consensus at the moment
 among both experimentalists and theorists about the {\it pairing} symmetry in KFe$_2$As$_2$. On the one hand,
non-phase-sensitive
  measurements on KFe$_{2}$As$_{2}$, such as thermal conductivity and Raman scattering, were interpreted
as evidence for a $d-$wave gap~\cite{louis,louis_2,raman}. On the other, laser ARPES reported full gap along the inner hole Fermi surface (FS), eight nodes along the middle FS, and negligible gap along the outer
 ($d_{xy}$) pocket~\cite{shin}. This was interpreted as evidence of $s-$wave pairing~\cite{shin,comm_n}.
 Specific heat data~\cite{sh} on KFe$_{2}$As$_{2}$ were also interpreted in favor of $s-$wave with multiple gaps.

Existing theoretical proposals for superconductivity in KFe$_{2}$As$_{2}$  explore the idea that the origin of the pairing in this system
 is the same as in FeSCs with hole and electron pockets, i.e., that the pairing is promoted by weak magnetic fluctuations.
This mechanism has been analyzed within RPA~\cite{laha,mkc} and within the renormalization group (RG)~\cite{frg_thomale}, and was recently re-discovered~\cite{osc}.
  The outcome is that, depending on parameters,
  spin fluctuations either favor $s^{+-}$ SC with the gap changing sign between the inner and the middle $d_{xz}/d_{yz}$ pockets~\cite{laha,mkc},
   or $d-$wave SC with the gap  predominantly residing on the outer $d_{xy}$ pocket~\cite{frg_thomale}

   Each scenario has a potential to explain superconductivity in KFe$_{2}$As$_{2}$, but
    the key shortcoming of both is that $s-$wave and the $d-$wave  attractions are very weak~\cite{mkc} because the mechanism is essentially of Kohn-Luttinger type~\cite{KL}.
Additionally, the $d-$wave pairing scenario yields the largest gap on the $d_{xy}$ pocket, which
is inconsistent with laser ARPES~\cite{shin}.

\begin{figure}[t]
\begin{center}
\includegraphics[width=0.49\columnwidth]{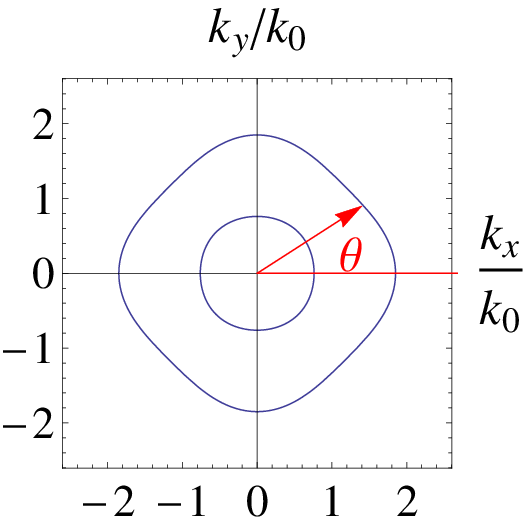} \includegraphics[width=0.45\columnwidth]{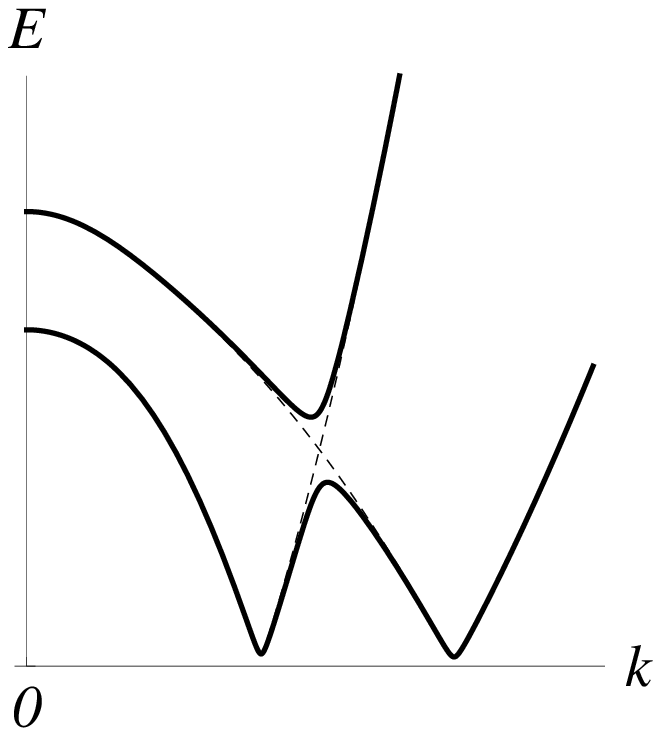}
\end{center}
\caption{Left panel:
 Illustrative Fermi surfaces (FS) for the $d_{xz}/d_{yz}$ hole pockets, where $k_0=\sqrt{2m\mu}$.
In the SC state, the pairing amplitude on the outer Fermi surface is $\Delta_+$ and on the inner $\Delta_-$.
Right panel:  Schematic quasiparticle dispersion in the superconducting state (solid black lines).  The gap away from the Fermi level is due to the $A_{2g}$ pairing and is present already without SO.
Once SO is included, the
 gaps on the FS appear. The dashed lines are approximations which capture the gaps on the FS only.
\label{fig:FS} }
\end{figure}

In this paper we propose a new mechanism for SC in KFe$_2$As$_2$ and other materials with only hole pockets. Consistent with laser ARPES\cite{shin}, we assume that the pairing involves mainly holes from $d_{xz}/d_{yz}$ pockets (see Fig.\ref{fig:FS}), and neglect the hole barrels near $(\pi,\pi)$ and the $d_{xy}$ pocket at $\Gamma$ where the observed pairing gap is much smaller. The pairing
in our theory arises from the combination of two factors:
sizable Hund's electron-electron interaction $J$ and sizable spin-orbit (SO) interaction $\lambda$. Specifically, we
argue that the system develops an $s-$wave SC
as soon as $J$ exceeds the inter-orbital Hubbard repulsion $U'$,
 {\it regardless} of the value of the intra-orbital Hubbard repulsion $U$.
The effective dimensionless coupling constant in the  $s-$wave pairing channel scales as
$N_0 (J-U') \left(\frac{\lambda}{\mu}\right)^2 $, where $N_0$ is the density of states and $\mu$ is the chemical potential.
That $J$ is substantial has been discussed in the context of ``Hund metal''~\cite{kotliar,leni}.
The magnitude of $\lambda$ is also quite sizable in FeSCs.
 ARPES measurements (Ref.\cite{borisenkoNatPhys2016})
 extracted $\lambda \sim 10-20$meV, comparable to $\mu$.

Without SO, the Cooper states at zero momentum can be classified according to their behavior separately under the crystal's point group operations and under spin SU(2) rotations.
As such, the on-site Hubbard-Hund interaction with positive $U$, $U'$, $J$ and $J'$
and $U > U', J, J'$ is
repulsive in the s-wave ($A_{1g}$) and $d-$wave ($B_{1g}$ and $B_{2g}$) spin singlet channels.
The interaction in the $A_{2g}$ spin-triplet channel, however, avoids $U$ and is $\frac{1}{2}(U'-J)$.
It
is attractive when $J > U'$ \cite{comm}.
 By itself, an attraction in the $A_{2g}$ channel does {\it not} necessarily lead to the Cooper
 instability because the pairing occurs between fermions from different bands and
  the pairing susceptibility
  is not logarithmically large at small temperature, $T$.  Besides, $A_{2g}$ pairing does not open gaps on the Fermi surfaces (see Fig.\ref{fig:FS}).
  The situation changes when $\lambda\neq 0$  because SO coupling
  mixes the $A_{1g}$ spin singlet and the $A_{2g}$ spin triplet pairs~\cite{CvetkovicVafek2013}.
 The pairing susceptibility in $A_{1g}$ channel diverges as  $\log{T}$
 at small $T$ because the order parameter contains
 fermion pairs from the same band. We argue that $s-$wave
 superconductivity emerges as soon as $J > U'$.
 Remarkably, this conclusion is unaffected by the presence of a much stronger $U$ despite the fact that the
 $U$ determines the repulsion in the $A_{1g}$ spin singlet channel.

The gaps on the two hole pockets are four-fold symmetric, but anisotropic. The solution of the self-consistency equations shows that the overall gap on the {\it larger} FS is smaller, in part, due to
 destructive interference between the $A_{1g}$ and the $A_{2g}$ components.
 For some range of parameters, the gap on this pocket has eight accidental nodes, as shown in the Fig.\ref{fig:gaps}.
  The relative magnitude of the $A_{1g}$ and the $A_{2g}$ components does not contain $\log{T}$, nevertheless, their ratio has a non-trivial temperature ($T$) dependence even at weak coupling.
This may lead to a possibility that such accidental nodes appear only below some $T < T_c$.

\begin{figure}[t]
\centering{}
\includegraphics[width=\columnwidth]{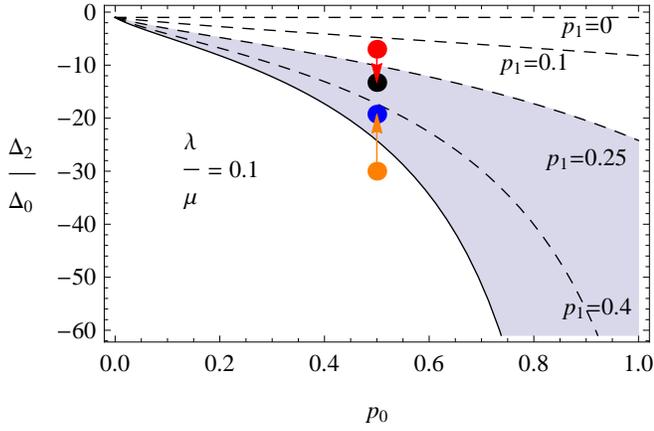}
\caption{The phase diagram at $T=0$ calculated at a fixed ratio of the SO coupling $\lambda$ to Fermi energy $\mu$.
Displayed are the boundaries of the nodal region, which depend on the ratio of $A_{2g}$ ($\Delta_2$) and $A_{1g}$ ($\Delta_0$) components of the pairing gap at $T=0$. They also depend on $p_0$ and $p_1$,
dimensionless parameters which enter into the angle dependence of the normal state band dispersion as in Eqs.(\ref{eq:band dispersion}) and (\ref{eq:Rtheta}).
The pairing amplitudes on the larger and the smaller Fermi surfaces are $\Delta_+ = \Delta_0  + (\lambda/|{\vec B}_{\bf k}|) \Delta_2$
  and  $\Delta_- = \Delta_0  - (\lambda/|{\vec B}_{\bf k}|) \Delta_2$, respectively; $2|{\vec B}_{\bf k}|$ is the energy of the band splitting (\ref{eq:band dispersion}). Shaded area marks the appearance of the accidental nodes in $\Delta_+$ for $p_1=0.25$.
  For a different value of $p_1$, the upper boundary of the shaded area shifts to the corresponding dashed line, while the lower boundary is $p_1$-independent.
  Below (above) the shaded region, the signs of $\Delta_+$ and $\Delta_-$ are opposite (same) and the pairing state can be viewed as $s^{+-}$ ($s^{++}$).
  Interestingly, numerical solutions of the self-consistency equations find that it is possible to start outside of the nodal region at $T_c$ (red and orange circles) and end up inside of it at $T=0$ (black and blue circles).
 \label{fig:phase diagram} }
\end{figure}

\begin{figure}[t]
\begin{center}
\includegraphics[width=0.49\columnwidth]{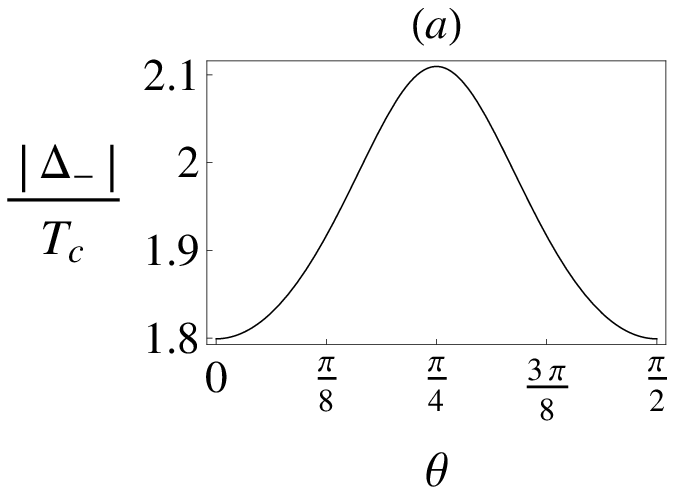} \includegraphics[width=0.49\columnwidth]{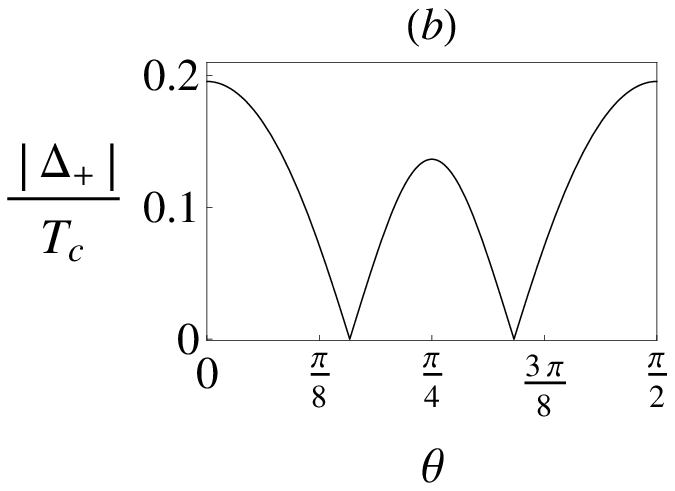}\\
\includegraphics[width=0.49\columnwidth]{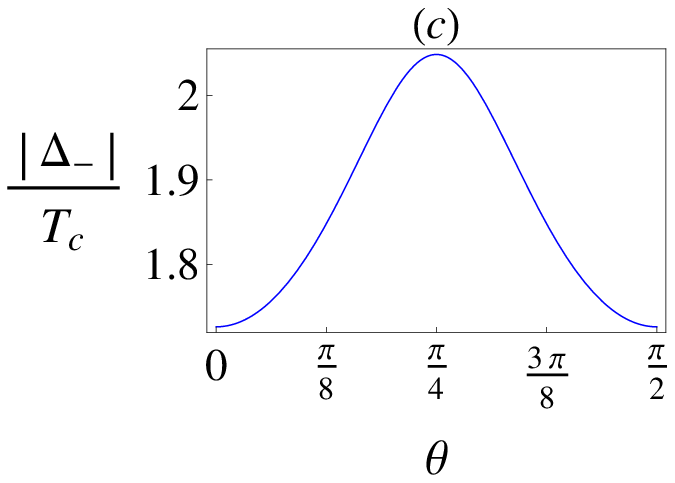} \includegraphics[width=0.49\columnwidth]{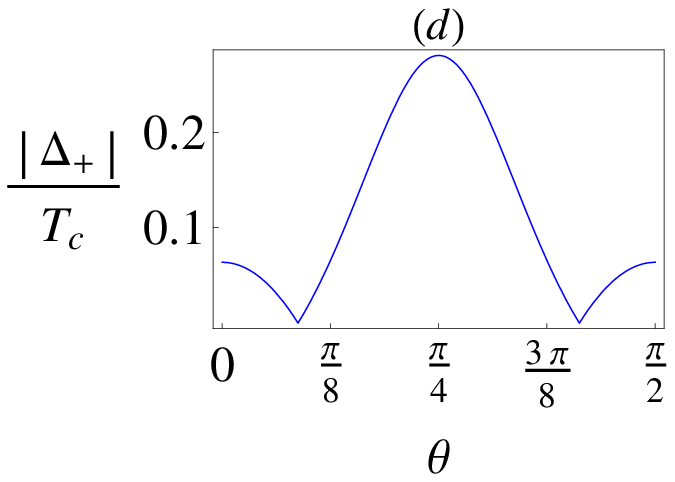}
\end{center}
  \caption{Angle dependence of the gap at $T=0$ on the inner (a) and the outer (b) hole Fermi surfaces (FS) for parameters corresponding to the
(black) end point of the down (red) arrow in Fig.\ref{fig:phase diagram}.
(c) and (d) show the same but for the parameters corresponding to the (blue) end point of the up (orange) arrow in Fig.\ref{fig:phase diagram}.
In both cases, there are eight nodal points on the outer FS.}\label{fig:gaps}
\end{figure}

Our results are summarized in Figs.\ref{fig:phase diagram} and \ref{fig:gaps}.  We argue below that they are consistent with several experimental findings
on K$_x$Ba$_{1-x}$Fe$_2$As$_2$ for $x \approx 1$.

{\it {\bf The model}}.~~~   We consider the itinerant model with two $\Gamma$-centered hole pockets made out of $d_{xz}$ and $d_{yz}$ orbitals (see Fig. \ref{fig:FS}).
The effective Hamiltonian
$\mathcal{H}=H_0+H_{int}$
 for the low-energy states near $\Gamma$  can be obtained, quite generally, using the method of invariants~\cite{CvetkovicVafek2013,rafael_oskar}, without the need to assume a particular microscopic model.
The  non-interacting part is
\begin{eqnarray}
H_0&=&\sum_{\bk}\sum_{\alpha,\beta=\uparrow,\downarrow} \psi^\dagger_{\bk,\alpha}\left(h_{\bk}\delta_{\alpha\beta}+h^{SO}s^z_{\alpha\beta}\right)\psi_{\bk,\beta},
\end{eqnarray}
where the doublet
 $\psi^\dagger_{\bk,\sigma}=\left(d^\dagger_{Yz,\sigma}(\bk),-d^\dagger_{Xz,\sigma}(\bk)\right)$,
 and
\begin{eqnarray}
h_{\bk}&=&\left(
\begin{array}{cc}
\mu-\frac{\bk^2}{2m}+bk_xk_y & c\left(k^2_x-k_y^2\right) \\
c\left(k^2_x-k_y^2\right) & \mu-\frac{\bk^2}{2m}-bk_xk_y
\end{array}
\right),\label{eq:hk}\\
h^{SO}&=&\lambda\left(
\begin{array}{cc}
0 & -i \\
i & 0
\end{array}
\right).\label{eq:hSO}
\end{eqnarray}
 The coefficients $\mu,m, b, c$, and the SO coupling $\lambda$ are material specific, but the forms of $h_{\bf k}$ and $h^{SO}$ are
 universal.

The 4-fermion interaction Hamiltonian  can
 also be written out in terms of the low energy doublet.
Assuming spin SU(2) symmetry and
 local interaction, we can express $H_{int}$ in real space as
\begin{eqnarray}\label{eq:Hint}
H_{int}&=&\sum_{j=0}^3\frac{g_j}{2} \int d^2\br :\!\psi^\dagger_{\sigma}(\br)\tau_j\psi_{\sigma}(\br)\psi^\dagger_{\sigma'}(\br)\tau_j\psi_{\sigma'}(\br)\!:,
\end{eqnarray}
where  $::$ implies normal ordering, the repeated spin indices $\sigma$, $\sigma'$ are summed over, $\tau_0=\bbone$ and the
three Pauli matrices $\tau_j$ act on the two components of the doublet.
The four couplings $g_j$ can be parameterized as
$g_0=\frac{1}{2}(U+U')$, $g_1=\frac{1}{2}(J+J')$, $g_2=\frac{1}{2}(J-J')$, and $g_3=\frac{1}{2}(U-U')$.
We emphasize that $g_i$'s include renormalizations from high energy modes
 and, in general, $U$, $U'$, $J$, and $J'$ are not the same as the {\it bare} Hubbard and Hund's interaction terms.

For $\lambda=0$, the pairing can be decomposed into spin singlet $A_{1g}$, $B_{1g}$, and  $B_{2g}$ channels, as well as the spin triplet $A_{2g}$.
The corresponding couplings are~\cite{CvetkovicVafek2013,ckf}
 $g_{A_{1g}} = {\tilde g}_0 = (U + J')/2$,  $g_{B_{1g}} = (U - J')/2$,  $g_{B_{2g}} = (U' + J)/2$, and  $g_{A_{2g}} = {\tilde g}_2 =\frac{1}{2}\left(g_0-g_1-g_2-g_3\right)= (U' - J)/2$.
The interactions in $A_{1g}$, $B_{1g}$, and  $B_{2g}$ channels are repulsive as the intra orbital Hubbard $U$ is the largest local interaction.
However the interaction in $A_{2g}$ channel is attractive if $J > U'$.
We assume this to hold.
 The $A_{2g}$ order parameter is
 \begin{eqnarray}\label{eq:Delta2}
 \Delta_{2} &=& \frac{1}{2}\tilde{g}_2\langle\psi^T_{\alpha}(\br)\tau_2(is^zs^y)_{\alpha\beta}\psi_{\beta}(\br)\rangle.
 \end{eqnarray}
Because $\tau_2$ is antisymmetric and $is^zs^y$ is symmetric, this order parameter is spin triplet.
For $\lambda =0$,
$\Delta_2$ in the band basis
 is composed entirely of fermions from different pockets.
  The susceptibility for such inter-pocket pairing  does not contain
the Cooper logarithm, and hence the attraction in  $A_{2g}$ channel
alone does not give rise to Cooper pairing, at least
at weak coupling.
However, in the presence of the SO interaction, an arbitrarily weak $A_{2g}$ attraction gives rise to a pairing instability, as we now show.

 {\it {\bf Role of SO coupling.}}  For $\lambda\neq 0$, the $A_{1g}$ and the $A_{2g}$ channel in Eq.(\ref{eq:Delta2}) mix\cite{CvetkovicVafek2013}.
Nevertheless, the $A-$channels and the $B-$channels remain decoupled.
We focus on the $A_{1g}$ channels because of the
attraction in $A_{2g}$.
Due to $A_{2g}/A_{1g}$ mixing, the order parameter $\Delta_2$ receives a contribution from fermions residing in the {\it same} band.  The corresponding normal state pairing susceptibility is logarithmically large at small $T$.
     There is a caveat, however -- the spin singlet $A_{1g}$ pairing component is strongly repulsive. Our goal is to analyze whether it prevents pairing
     when $\tilde{g}_{2}<0$.
To this end, we also introduce the conventional spin singlet $A_{1g}$ order parameter,
\begin{eqnarray}\label{eq:Delta0}
\Delta_{0} &=& \frac{1}{2}\tilde{g}_0\langle\psi^T_{\alpha}(\br)\bbone(-is^y)_{\alpha\beta}\psi_{\beta}(\br)\rangle,
\end{eqnarray}
and  obtain the set of two coupled
 equations
 for $\Delta_2$ and $\Delta_0$ (Ref. \cite{sm}).
At $T_c$, we have for ${\tilde g}_2 <0$ and ${\tilde g}_0>0$
\begin{eqnarray}\label{eq:self consistency at Tc0}
-\frac{\Delta_0}{\tilde g_0}&=&\sum_{\rho=\pm}\int\frac{d^2 \bk}{(2\pi)^2}\frac{\tanh\frac{\xi_{\rho}}{2T_c}}{2\xi_\rho}
\left(\Delta_0+\rho\Delta_2\frac{\lambda}{|\vec{B}_\bk|}\right),\\
-\frac{\Delta_2}{\tilde g_2}&=&\sum_{\rho=\pm}
\int\frac{d^2 \bk}{(2\pi)^2}\frac{1}{2\xi_\rho}\tanh\frac{\xi_{\rho}}{2T_c}\times\label{eq:self consistency at Tc2}
\\
&\times&\left(\Delta_2\left(\frac{\lambda^2}{{\vec B}^2_\bk}+\frac{\xi_\rho}{A_\bk}\left(1-\frac{\lambda^2}{{\vec B}^2_\bk}\right)\right)+\rho\Delta_0\frac{\lambda}{|\vec{B}_\bk|}\right)\nonumber,
\end{eqnarray}
where the normal state
band
dispersion has the form
\begin{eqnarray}\label{eq:band dispersion}
\xi_{\pm}&=&A_{\bk}\pm |\vec{B}_\bk|=\mu-\frac{\bk^2}{2m}\pm\sqrt{R_\theta \frac{\bk^4}{4m^2}+\lambda^2}.
\end{eqnarray}
The angular anisotropy in momentum space enters via $0<R_\theta<1$, and is determined by the coefficients $b$ and $c$ in Eq.(\ref{eq:hk}).
We express it as
\begin{eqnarray}\label{eq:Rtheta}
R_\theta = p_0 \left(\frac{1}{2} + p_1 + \left(\frac{1}{2} - p_1\right) \cos4\theta\right),
\end{eqnarray}
with $p_0=4m^2c^2$ and $p_1=b^2/(8c^2)$. Without loss of generality, $0<p_0<1$ and $0<p_1<\frac{1}{2}$.
The Fermi surfaces shown in Fig.(\ref{fig:FS}) correspond to $p_0=0.5$, $p_1=0.4$, and $\lambda/\mu=0.1$.
Eqs. (\ref{eq:self consistency at Tc0}-\ref{eq:self consistency at Tc2}) have the form
\begin{eqnarray}\label{eq:Deltas at Tc}
\left(\begin{array}{cc}
-\frac{1}{\tilde g_0}-\chi_{00}(T_c) & -\chi_{02}(T_c) \\
-\chi_{02}(T_c) & -\frac{1}{\tilde g_2}-\chi_{22}(T_c)
\end{array}\right)
\left(\begin{array}{c}
\Delta_0(T_c) \\ \Delta_2(T_c)
\end{array}\right)=0.\nonumber\\
\end{eqnarray}
Therefore, $T_c$ is determined from requiring that the determinant vanishes
\begin{eqnarray}\label{eq:Tc}
-\frac{1}{\tilde g_2}+\frac{\chi^2_{02}(T_c)}{\frac{1}{\tilde g_0}+\chi_{00}(T_c)}&=&\chi_{22}(T_c).
\end{eqnarray}
Brief inspection of (\ref{eq:self consistency at Tc0}-\ref{eq:self consistency at Tc2}) reveals that $\chi_{00}$ and $\chi_{22}$
 scale as
  $\sim\ln{\frac{1}{T}}$.
  On the other hand, $\chi_{02}(T)$ remains finite
   due to an {\it exact cancellation} of two such logs.
For $\mu\gg T_c$, we find
\begin{equation}
\chi_{02}(T_c)=\frac{m}{2\pi}\frac{\lambda}{\mu}\int_0^{2\pi}\frac{d\theta}{2\pi}
\frac{\tanh^{-1}\sqrt{R_\theta+(1-R_\theta)\frac{\lambda^2}{\mu^2}}}
{\sqrt{R_\theta+(1-R_\theta)\frac{\lambda^2}{\mu^2}}},
\label{f_4}
\end{equation}
where $\tanh^{-1}x=\frac{1}{2}\ln\frac{1+x}{1-x}$.
As a result, $T_c$ is finite regardless of how weak is the attractive coupling, $\tilde{g}_2<0$, and how strong is the repulsive coupling $\tilde{g}_0>0$.
Moreover, $\chi_{02}(T_c) \lambda/\mu$ is positive.
From the gap equations we then find that
$\Delta_0(T_c)= - \mathcal{C}\Delta_2(T_c) \lambda/\mu$, where $\mathcal{C}>0$.
The gaps on the two pockets are
\begin{eqnarray}\label{eq:DeltaPM}
\Delta_\pm=\Delta_0 \pm \frac{\lambda}{|{\vec B}_{\bf k}|} \Delta_2,
\end{eqnarray}
where $\Delta_+$ is on the larger and $\Delta_-$ is on on the smaller pocket.
Analyzing the forms of these gaps, we find that
(i) $|\Delta_+|$ is reduced relative to $|\Delta_-|$, (ii) the gaps are four-fold symmetric, but anisotropic, and
(iii) for small $|{\tilde g}_2|$,
$\Delta_0$ is
small compared to $\Delta_2$, forcing opposite signs of $\Delta_+$ and $\Delta_-$, i.e. $s^{+-}$
gap structure.

{\it {\bf Below $T_c$}.}~~~~
The mean field equations below $T_c$ are non-linear in $\Delta_0(T)$ and $\Delta_2(T)$. We eliminate the couplings $\tilde{g}_0$ and $\tilde{g}_2$ by expressing $\Delta_0$ and $\Delta_2$ in units of $T_c$. Solving the non-linear set
 we obtain
 $\Delta_{0,2}(T)/T_c$ and the ratio $ K(T) = \Delta_0 (T)/\Delta_2 (T)$ in terms of the same ratio at $T_c$.
 In a general case, when the cross term $\chi_{0,2}$  is non-logarithmic, $K(T)$ remains the same as at $T_c$, at least at weak coupling.
 In our case, the situation is different because a finite $\chi_{02} (T)$ is due to subtle cancellation of the logs, and leftover terms are $T$-dependent.
 In the limit of
$K(T_c) \ll 1$
 we found analytically $K(T=0) = K(T_c) (1 + \mathcal{A})$, where $\mathcal{A}>0$ (Ref. \cite{sm}).
This also holds in the numerical solution of the mean-field equation, as indicated by the lower arrow in the Fig.\ref{fig:phase diagram}.

The numerical and analytical considerations show that the gap may  have accidental nodes.
The numerical solutions of the gap equations are shown in the Fig. \ref{fig:gaps}.
We see that, indeed, in some range of parameters, the gap on the larger hole pocket has eight accidental nodes.
Interestingly, as shown in the Fig.\ref{fig:phase diagram}, we also found that over some range of parameters the nodes are absent at $T_c$, but appear at $T=0$.

{\it {\bf Comparison with experiments.}}
Our results are consistent with several experimental findings
on K$_x$Ba$_{1-x}$Fe$_2$As$_2$ for $x \approx 1$.
 Namely, (i)  a larger gap on the inner hole pocket at $\Gamma$, with no nodes,
(ii) a smaller gap magnitude and the appearance of the accidental nodes on the larger $d_{xz}/d_{yz}$ pocket (middle pocket at $\Gamma$), and (iii)  angular correlation of the gap maxima on the two FSs  are all  consistent with the ARPES results~\cite{shin}.
 The presence of the gap nodes is
 consistent with thermal conductivity and Raman scattering measurements~\cite{louis,louis_2,raman},
   and the near-absence of the gap on the $d_{xy}$ pocket is consistent with ARPES~\cite{shin} and specific heat measurements~\cite{sh}.
   We also analyzed the temperature dependence of the
   the spin susceptibility $\chi (T)$ by adding a Zeeman coupling to $\mathcal{H}$.
   We found that $\chi (T)$  decreases below $T_c$ for {\it any} orientation of the external magnetic field, even if $\Delta_0$ is negligible compared to $\Delta_2$.
  This result is non-trivial because for $\lambda=0$ the pairing
   was in $A_{2g}$ spin-triplet channel, and $\chi(T)$
    was
     {\it not} suppressed below $T_c$ when the magnetic field
   is perpendicular to the triplet ${\bf d}$-vector.
   The decrease of $\chi (T)$ for any orientation of the magnetic field is consistent with the Knight shift measurements in KFe$_2$As$_2$ (Ref. \cite{Fukazawa2011}).
      Finally, from Eq.(\ref{eq:self consistency at Tc2}) we readily see that the prefactor of the Cooper logarithm in $\chi_{22}(T_c)$ contains a factor of $\lambda^2/\mu^2$. Therefore $T_c$ {\it increases} as $\mu$ decreases,
  for fixed ${\tilde g}_{0,2}$  and fixed $\lambda$.  When $T_c \ll \mu$, we found, to logarithmic accuracy,
  \begin{eqnarray}
\frac{T_c}{\mu}\sim \exp\left({-\sqrt{\left(1+p_0\frac{\mu^2}{\lambda^2}\right)\left(1+2p_0p_1\frac{\mu^2}{\lambda^2}\right)}\frac{\pi}{m|\tilde g_2|}}\right).
\end{eqnarray}
  The increase of $T_c$ with decreasing $x$ is consistent with the $x$ dependence of $T_c$
   in  K$_x$Ba$_{1-x}$Fe$_2$As$_2$ at $x \leq 1$.  At smaller $x$, electron pockets appear, and $s-$wave pairing may become caused by interaction between fermions near hole and electron pockets.

 {\it {\bf Conclusions.}}~~~ In this paper we presented a novel mechanism of $s-$wave pairing in FeSC, which involves fermions near $d_{xz}/d_{yz}$ hole pockets.
   When the renormalized Hund's interaction $J$ exceeds the renormalized inter-orbital Hubbard repulsion $U'$, the interaction in $A_{2g}$ channel is attractive. In the absence of SO coupling, this attraction would potentially give rise to spin-triplet superconductivity, but only when the attractive coupling exceeds a certain threshold. We argued that at a non-zero SO coupling, the same interaction gives an attraction in the $s$-wave channel, where the pairing condensate involves fermions from the same band and superconductivity emerges at an arbitrarily weak attraction.  We demonstrated that $T_c$ is only weakly affected by the large inter-orbital repulsion $U$ in the $A_{1g}$ channel, despite the fact that the SO coupling mixes the $A_{2g}$ and the $A_{1g}$ components.  The gap functions are four-fold symmetric, but anisotropic, particularly on the larger FS, where over some range of parameters the gap has accidental nodes.   Our results are consistent with ARPES and other experiments on strongly hole doped  K$_x$Ba$_{1-x}$Fe$_2$As$_2$.

We thank E. Berg, P. Hirschfeld, R. Fernandes, M. Khodas, and  J. Schmalian for useful discussions.
 OV was supported by NSF DMR-1506756.  AVC was supported
by the Office of Basic Energy Sciences, U.S. Department of Energy,
under awards DE-SC0014402.
The authors thank the
Aspen Center for Physics,
where part of this work was performed, for its hospitality. ACP is supported by NSF grant PHY-1066293.

\section{Supplementary Material}

\subsection{Pairing in the orbital and band representations}

In this Section we discuss how one can understand the results of random-phase approximation (RPA) and functional renormalization group (fRG) analysis of the pairing in systems with only hole pockets.
 Spin fluctuations generally require the presence of both hole and electron pockets and are weak in systems with only one type of pockets. Yet, both
  RPA and fRG calculations showed that even a weak renormalization of Hubbard and Hund interactions gives rise to an attraction in s-wave and d-wave channels.
   The attractive interaction is weak and the corresponding $T_c$ is truly small and is very likely much smaller than $s-$wave $T_c$ that we obtained in this paper.
    Nevertheless, as a matter of principle, the attraction does appear in numerical calculations,  and below we  show how one can understand analytically why it emerges.
     Another goal of our discussion is to clarify the interplay between $s-$wave and $d-$wave order parameters in the orbital and the band basis.

     We begin by noticing that in a system with a local Hubbard and Hund interaction, it is natural to classify the pairing states in the orbital basis because  pairing interaction in this basis decouples between different channels.   For the same model as in the bulk of the paper (i.e., the model of fermions on $d_{xz}$ and $d_{yz}$ orbitals near the $\Gamma$ point),  the order parameters in the $s$-wave ($A_{1g}$) and d-wave ($B_{1g}$) channels are
     \be
     &&\Delta^{orb}_{A_{1g}} = d^\dagger_{xz,\uparrow}  d^\dagger_{xz,\downarrow} +  d^\dagger_{yz,\uparrow}  d^\dagger_{yz,\downarrow}, \nonumber \\
     &&\Delta^{orb}_{B_{1g}} = d^\dagger_{xz,\uparrow}  d^\dagger_{xz,\downarrow} -  d^\dagger_{yz,\uparrow}  d^\dagger_{yz,\downarrow}.
     \label{s_0}
     \ee
The Hubbard-Hund local Hamiltonian contains intra-pocket and inter-pockets Hubbard terms ($U$ and $U'$ terms, respectively), the Hund exchange $J$ term and the Hund pair hopping term $J'$. Out of these four terms, Hubbard $U$ and Hund $J'$ terms contribute to the pairing Hamiltonian at the mean-field level (i.e., without renormalizations).
 In momentum space, the pairing Hamiltonian takes the form
\begin{widetext}
\begin{align}
H_{orb}= & U \sum_{k,p}  \left (d_{xz,k,\uparrow}^{\dag}d_{xz, -k,\downarrow}^\dag d_{xz,p, \downarrow} d_{xz, -p, \uparrow}
  +  d_{yz,k,\uparrow}^{\dag}d_{yz, -k,\downarrow}^\dag d_{yz,p, \downarrow} d_{yz, -p, \uparrow}  + h.c\right) \nonumber \\
& +  J' \sum_{k,p}  \left (d_{xz,k,\uparrow}^{\dag}d_{xz, -k,\downarrow}^\dag d_{yz,p, \downarrow} d_{yz, -p, \uparrow}
  +  d_{yz,k,\uparrow}^{\dag}d_{yz, -k,\downarrow}^\dag d_{xz,p, \downarrow} d_{xz, -p, \uparrow}  + h.c\right)
\label{s_01}
\end{align}
\end{widetext}
This Hamiltonian can be equivalently re-written as
\beq
H_{orb}= \frac{U+J'}{2} |\Delta^{orb}_{A_{1g}}| + \frac{U-J'}{2} |\Delta^{orb}_{B_{1g}}|
\label{s_02}
\eeq
The corresponding eigenvalues $(U + J')/2$  for $A_{1g}$ and  $(U - J')/2$  for $B_{1g}$ are both negative as long as $U > J'$.  The corrections from dressing the interaction by particle-hole bubbles cannot change the sign of the interaction, at least at weak coupling and away from a collective instability.

      We now switch gears and consider $s-$wave and $d-$wave ($A_{1g}$ and $B_{1g}$) order parameters in the band basis.
        For simplicity, we assume that hole pockets are circular. An extension to non-circular, but still $C_4-$symmetric pockets is straightforward and just complicates the formulas without changing the results.

       Let's denote band fermions as $d_1$ and $d_2$.   A simple experimentation shows that there are four possible order parameters
      \be
      &&\Delta^{b}_{s^{++}} (k) =  d^\dagger_{1,k,\uparrow}  d^\dagger_{1,-k,\downarrow} +  d^\dagger_{2,k,\uparrow}  d^\dagger_{2,-k,\downarrow} \nonumber\\
      &&\Delta^{b}_{s^{+-}} (k) =  d^\dagger_{1,k,\uparrow}  d^\dagger_{1,-k} -  d^\dagger_{2,k}  d^\dagger_{2,-k,\downarrow} \nonumber\\
      &&\Delta^{b}_{d^{++}} (k) =  \left(d^\dagger_{1,k,\uparrow}  d^\dagger_{1,-k} +  d^\dagger_{2,k,\uparrow}  d^\dagger_{2,-k,\downarrow}\right) \cos{2 \theta_k} \nonumber\\
       &&\Delta^{b}_{d^{+-}} (k) =  \left(d^\dagger_{1,k,\uparrow}  d^\dagger_{1,-k} -  d^\dagger_{2,k,\uparrow}  d^\dagger_{2,-k,\downarrow}\right) \cos{2 \theta_k}
       \label{s_4}
\ee
where $\theta$ is the angle along each of the  Fermi surfaces, counted from, say, $x-$axis.
The first two order parameters have $s-$wave symmetry -- a conventional $s^{++}$  and $s^{+-}$, which changes sign between the two bands. The other two have $d-$wave symmetry, again with or without additional sign change between the two pockets ($d^{+-}$ and $d^{++}$, respectively).

Clearly, there are more options in the band basis than in the orbital basis.  To understand why there is a (potential) discrepancy, we convert the local interaction from the orbital to the band basis.  For circular pockets, the transformation from orbital to band basis is just a rotation:
\be
&&d_{xz} (k,\sigma) = d_1 (k,\sigma) \cos{\theta} + d_2 (k,\sigma) \sin{\theta}, \nonumber \\
&& d_{yz} (k,\sigma) = d_2 (k,\sigma) \cos{\theta} - d_1 (k,\sigma) \sin{\theta};
\label{s_6}
\ee
Ttransforming the interaction Hamiltonian, Eq. (\ref{s_01}) from orbital to band basis,  we obtain
\begin{widetext}
\begin{align}
H_{b}= &  \sum_{k,p} \left(\frac{U+J'}{2} + \frac{U-J'}{2}\cos{\theta_k} \cos{\theta_p} \right) \left(d_{1,k,\uparrow}^{\dag}d_{1, -k,\downarrow}^\dag d_{1,p, \downarrow} d_{1, -p, \uparrow} + d_{2,k,\uparrow}^{\dag}d_{2, -k,\downarrow}^\dag d_{2,p, \downarrow} d_{2, -p, \uparrow}  + h.c\right) \nonumber \\
& +  \sum_{k,p} \left(\frac{U+J'}{2} - \frac{U-J'}{2}\cos{\theta_k} \cos{\theta_p} \right) \left(d_{1,k,\uparrow}^{\dag}d_{1, -k,\downarrow}^\dag d_{2,p, \downarrow} d_{2, -p, \uparrow} + d_{2,k,\uparrow}^{\dag}d_{2, -k,\downarrow}^\dag d_{1,p, \downarrow} d_{1, -p, \uparrow}  + h.c\right)
\label{s_1}
\end{align}
\end{widetext}
This Hamiltonian can be equivalently re-written as
\beq
H_b = \frac{U+J'}{2} \sum_k |\Delta^{b}_{s^{++}} (k)|^2 + \frac{U-J'}{2} \sum_k |\Delta^{b}_{d^{+-}} (k)|^2
\label{s_3}
\eeq
which is the same as Eq. (\ref{s_02}).
We see that, as expected, only two gap functions are present, one in $A_{1g}$ channel and the other in $B_{1g}$ channel.  The two other order parameters, $s^{+-}$ and $d^{++}$, do not appear in the Hamiltonian, i.e., the corresponding couplings are strictly zero.

Let's now continue with the band basis analysis and include the effect of renormalization of the pairing interaction by particle-hole bubbles.  This can be second-order renormalization by a single bubble (Kohn-Luttinger effect) or it may include RPA series of particle-hole bubbles. In the latter case the effect of RPA summation
 can be re-expressed as due to collective spin fluctuations.  Spin fluctuations are rather weak in KFe$_2$As$_2$, so most likely the dominant renormalization at not too strong coupling comes from a single particle-hole bubble.  The renormalization affects differently the prefactors in different terms in Eq. (\ref{s_1}).  In a generic case, $H_b$ changes to
 \begin{widetext}
 \begin{align}
&H_{b}= \nonumber \\
 &  \sum_{k,p} \left(U_{11} + {\bar U}_{11} \cos{\theta_k} \cos{\theta_p} \right) \left(d_{1,k,\uparrow}^{\dag}d_{1, -k,\downarrow}^\dag d_{1,p, \downarrow} d_{1, -p, \uparrow} + h.c\right)
  + \left(U_{22} + {\bar U}_{22} \cos{\theta_k} \cos{\theta_p} \right)  \left( d_{2,k,\uparrow}^{\dag}d_{2, -k,\downarrow}^\dag d_{2,p, \downarrow} d_{2, -p, \uparrow}  + h.c \right) \nonumber \\
& +  \sum_{k,p} \left(U_{12} - {\bar U}_{12}\cos{\theta_k} \cos{\theta_p} \right)  \left(d_{1,k,\uparrow}^{\dag}d_{1, -k,\downarrow}^\dag d_{2,p, \downarrow} d_{2, -p, \uparrow} + d_{2,k,\uparrow}^{\dag}d_{2, -k,\downarrow}^\dag d_{1,p, \downarrow} d_{1, -p, \uparrow}  + h.c\right)
\label{s_2}
\end{align}
\end{widetext}
where in the absence of renormalizations $U_{11} = U_{22} = U_{12} = (U+J')/2$ and ${\bar U}_{11} = {\bar U}_{22} = {\bar U}_{12} = (U-J')/2$.
To make presentation easier to follow, we assume that $U_{11}= U_{22}$ and ${\bar U}_{11} = {\bar U}_{22}$ even after renormalization,  but keep $U_{11} \neq U_{12}$  and ${\bar U}_{11} \neq {\bar U}_{12}$.  Decomposing $H_b$ into contributions with different order parameters, like we did in going from (\ref{s_1}) to (\ref{s_3}) we immediately find that
  $H_b$ now contains contributions with all four order parameters from (\ref{s_4}):
\begin{widetext}
\be
&&H_b = \frac{U_{11} + U_{12}}{2} \sum_k |\Delta^{b}_{s^{++}} (k)|^2 + \frac{U_{11} - U_{12}}{2} \sum_k |\Delta^{b}_{s^{+-}} (k)|^2  \nonumber \\
&& + \frac{{\bar U}_{11} + {\bar U}_{12}}{2} \sum_k |\Delta^{b}_{d^{+-}} (k)|^2 + \frac{{\bar U}_{11} - {\bar U}_{12}}{2} \sum_k |\Delta^{b}_{d^{++}} (k)|^2 \nonumber \\
\label{s_5}
\ee
\end{widetext}
For $A_{1g}$ channel this is nothing but a well-known generation of $s^{+-}$ interaction by a renormalization which makes  intra-pocket repulsion different from inter-pocket repulsion.  When  renormalization makes $U_{12}$ larger than $U_{11}$, the system develops an attractive interaction in $s^{+-}$ channel, and arbitrary weak attraction already gives rise to a BCS instability in $s^{+-}$ channel, despite strong repulsion in $s^{++}$  channel (the situation becomes more complex beyond BCS as $s^{++}$ and $s^{+-}$ order
 parameters obviously belong to the same $A_{1g}$ representation and hence in general do not decouple).  The weak attraction in $s^{+-}$ channel has been found in Ref. \cite{mkc} using spin-fluctuation calculations and band structure for KFe$_2$As$_2$ and cited there as a potential reason for $s^{+-}$ pairing in this material.  Note, however, that the same mechanism may give rise to a weak d-wave ($d^{++}$)  pairing, if renormalized ${\bar U}_{12}$ exceeds ${\bar U}_{11}$.

We now go back to orbital basis and check how the interaction and the gap structure in $s^{+-}$ and $d^{++}$ channels looks like there.  The gap structure is easily obtained by
 inverting the transformation  (\ref{s_6}):
 \be
&&d_{1} (k,\sigma) = d_{xz} (k,\sigma) \cos{\theta} - d_{yz} (k,\sigma) \sin{\theta}, \nonumber \\
&& d_{2} (k,\sigma) = d_{yz} (k,\sigma) \cos{\theta} + d_{xz} (k,\sigma) \sin{\theta}.
\label{s_7}
\ee
Substituting this transformation into (\ref{s_4}) we indeed recover Eq. (\ref{s_0}) for $s^{++}$ and $d^{+-}$ order parameters (labeled $A_{1g}$ and $B_{1g}$ in (\ref{s_0}), modulo an additional $(1 +\cos{4 \theta})$ factor in $\Delta^{orb}_{B_{1g}}$. The other two order parameters
in the orbital representation are
\be
     &&\Delta^{orb}_{s^{+-}} = \left(d^\dagger_{xz,\uparrow}  d^\dagger_{xz,\downarrow} -  d^\dagger_{yz,\uparrow}  d^\dagger_{yz,\downarrow}\right) \cos{2\theta}, \nonumber \\
     &&\Delta^{orb}_{d^{++}} = \left(d^\dagger_{xz,\uparrow}  d^\dagger_{xz,\downarrow} +  d^\dagger_{yz,\uparrow}  d^\dagger_{yz,\downarrow}\right) \cos{2\theta}
 \label{s_8}
     \ee
 We see that $s^{+-}$ order parameter in orbital representation is a product of $B_{1g}$ order parameter from (\ref{s_0}) and $d-$wave form factor $\cos{2\theta}$. The product is indeed $C_4$ symmetric, as $s-$wave order parameter should be.

 To see how these new order parameter emerge if we solve for the pairing in the orbital basis, without moving back and forth  orbital basis, we re-express the renormalized interaction $H_b$ in the orbital basis.  Substituting (\ref{s_7}) into (\ref{s_5}) we find that the renormalization brings in additional pairing terms to originally local Hubbard-Hund interaction, in the form
\begin{widetext}
 \be
&&\delta H_{orb} = \lambda_1 \sum_{k,p} \cos{2 \theta_k} \cos{2\theta_p}  \left (d_{xz,k,\uparrow}^{\dag}d_{xz, -k,\downarrow}^\dag d_{xz,p, \downarrow} d_{xz, -p, \uparrow}
  +  d_{yz,k,\uparrow}^{\dag}d_{yz, -k,\downarrow}^\dag d_{yz,p, \downarrow} d_{yz, -p, \uparrow}  + h.c\right) \nonumber \\
&& +  \lambda_2 \sum_{k,p} \cos{2 \theta_k} \cos{2\theta_p}  \left (d_{xz,k,\uparrow}^{\dag}d_{xz, -k,\downarrow}^\dag d_{yz,p, \downarrow} d_{yz, -p, \uparrow}
  +  d_{yz,k,\uparrow}^{\dag}d_{yz, -k,\downarrow}^\dag d_{xz,p, \downarrow} d_{xz, -p, \uparrow}  + h.c\right)
\label{s_9}
\ee
\end{widetext}
where
\be
&&\lambda_1 = \frac{U_{11} - U_{12}}{2} + \frac{{\bar U}_{11} - {\bar U}_{12}}{2}, \nonumber \\
&&\lambda_2 = -\frac{U_{11} - U_{12}}{2} + \frac{{\bar U}_{11} - {\bar U}_{12}}{2},
\ee
We see that additional terms in $H_{orb}$ make the interaction in the orbital basis non-local and also dependent on the direction in the momentum space.
The interaction term $\delta H_{orb}$  looks like a $d-$wave term because of $\cos{2 \theta_k} \cos{2\theta_p}$ factors. However, $(U_{11} - U_{12})/2$ terms in $\lambda_1$ and $\lambda_2$ are also of different sign, and this additional sign change makes the corresponding part of $\delta H_{orb}$ $C_4$ symmetric.  Solving for the pairing right in the orbital basis we indeed obtain that $C_4$-symmetric part of $\delta H_{orb}$ gives rise to $s^{+-}$  pairing in $U_{12} > U_{11}$, while $C_4$ anti-symmetric part of  $\delta H_{orb}$ gives rise to $d^{++}$  pairing in ${\tilde U}_{12} > {\tilde U}_{11}$.

\subsection{Gap equations}

The effective BCS Hamiltonian for coupled spin singlet $A_{1g}$ and spin triplet $A_{2g}$ order parameters is
\begin{eqnarray}
\mathcal{H}\approx \mathcal{H}_{BdG}- L^2 \left(\frac{|\Delta_{0}|^2}{\tilde{g}_0}+\frac{|\Delta_{2}|^2}{\tilde{g}_2}\right),
\end{eqnarray}
where $L^2$ is the area of the system and
\begin{eqnarray}
\mathcal{H}_{BdG}&=&\sum_{\bk}\Psi^\dagger_\bk H_{BdG}(\bk) \Psi_\bk,
\label{f_1}
\end{eqnarray}
where in the Nambu notation
$\Psi^\dagger_\bk =\left(\psi^\dagger_{\bk\uparrow},\psi^T_{-\bk\downarrow}\right)$
and
\begin{eqnarray}\label{eq:Hbdg}
H_{BdG}(\bk) &=& \left(
\begin{array}{cc}
  h_\bk+\lambda \tau_2 & \bbone \Delta_0+\tau_2 \Delta_2 \\
  \bbone \Delta^*_0+\tau_2 \Delta^*_2 & -h_\bk-\lambda \tau_2
\end{array}
\right).
\label{f_2}
\end{eqnarray}
We used the fact that $h_\bk=h^T_{-\bk}$.

The linearized mean-field self-consistency equations at $T=T_c$ are presented in the main text.
 The equations below $T_c$ are $(k_B=1)$
\begin{eqnarray}\label{eq:self consistency0}
-\frac{\Delta_0}{\tilde g_0}&=&\sum_{p=\pm}\int\frac{d^2 \bk}{(2\pi)^2}\frac{\tanh\frac{E_{p}}{2T}}{2E_p}
\left(\Delta_0+4\Delta_2\frac{A_\bk\lambda+\Delta_0\Delta_2}{E^2_{p}-E^2_{-p}}\right),\nonumber\\
\\
-\frac{\Delta_2}{\tilde g_2}&=&\sum_{p=\pm}
\int\frac{d^2 \bk}{(2\pi)^2}\frac{\tanh\frac{E_{p}}{2T}}{2E_p}\times\nonumber\\
&\times&\left(\Delta_2+4\Delta_2\frac{\vec{B}^2_\bk-\lambda^2}{E^2_p-E^2_{-p}}+4\Delta_0\frac{A_\bk\lambda+\Delta_0\Delta_2}{E^2_p-E^2_{-p}}\right).\nonumber\\
\label{eq:self consistency2}\end{eqnarray}
The coefficients $A_\bk$ and $\vec{B}_\bk$ are related to the parameters of the  non-interacting Hamiltonian
\begin{eqnarray}
H_0&=&\sum_{\bk}\sum_{\alpha,\beta=\uparrow,\downarrow} \psi^\dagger_{\bk,\alpha}\left(h_{\bk}\delta_{\alpha\beta}+h^{SO}s^z_{\alpha\beta}\right)\psi_{\bk,\beta},
\end{eqnarray}
where
 $\psi^\dagger_{\bk,\sigma}=\left(d^\dagger_{Yz,\sigma}(\bk),-d^\dagger_{Xz,\sigma}(\bk)\right)$,
 and
\begin{eqnarray}
h_{\bk}&=&\left(
\begin{array}{cc}
\mu-\frac{\bk^2}{2m}+bk_xk_y & c\left(k^2_x-k_y^2\right) \\
c\left(k^2_x-k_y^2\right) & \mu-\frac{\bk^2}{2m}-bk_xk_y
\end{array}
\right),\label{eq:hk}\\
h^{SO}&=&\lambda\left(
\begin{array}{cc}
0 & -i \\
i & 0
\end{array}
\right). \label{eq:hSO}
\end{eqnarray}
The relation are
\begin{eqnarray}
h_\bk+\lambda \tau_2 &=& A_\bk \bbone +\vec{B}_\bk\cdot \vec{\tau}.
\end{eqnarray}

The two branches of the Bogoliubov quasiparticle dispersion are
\begin{eqnarray}\label{eq:dispersion_1}
E_{\pm}&=&\left(A^2_\bk+{\vec B}^2_\bk+\Delta^2_0+\Delta^2_2\right.\\
&\pm&\left.2\sqrt{A^2_\bk \vec{B}^2_\bk +\Delta^2_0\Delta^2_2+2\Delta_0\Delta_2A_\bk\lambda+\Delta^2_2(\vec{B}^2_\bk-\lambda^2)
}\right)^{\frac{1}{2}}.\nonumber
\end{eqnarray}
We show it schematically by the bold lines in the right panel of Fig. 1 in the main text.

To understand $H_{BdG}(\bk)$ in the band basis, we perform a unitary operation which diagonalizes $h_\bk+\lambda \tau_2$.
We find
\begin{widetext}
\begin{eqnarray}\label{eq:band BdG_1}
\mathcal{U}^\dagger H_{BdG}(\bk) \mathcal U=\left(\begin{array}{cccc}
\xi_+ & 0 & \Delta_0+\frac{\lambda}{|\vec{B}_\bk|}\Delta_2 & -i\Delta_2 \sqrt{1-\frac{\lambda^2}{\vec{B}^2_\bk}}\\
0 & \xi- & i\Delta_2 \sqrt{1-\frac{\lambda^2}{\vec{B}^2_\bk}} &  \Delta_0-\frac{\lambda}{|\vec{B}_\bk|}\Delta_2\\
\Delta_0+\frac{\lambda}{|\vec{B}_\bk|}\Delta_2 & -i\Delta_2 \sqrt{1-\frac{\lambda^2}{\vec{B}^2_\bk}} & -\xi_+ & 0 \\
i\Delta_2 \sqrt{1-\frac{\lambda^2}{\vec{B}^2_\bk}} & \Delta_0-\frac{\lambda}{|\vec{B}_\bk|}\Delta_2 & 0 &-\xi_-
\end{array}
\right).
\end{eqnarray}
\end{widetext}
At low energy, we can ignore the off-diagonal terms  in the pairing blocks. The Bogoliubov quasiparticle dispersion can be readily read off and approximated very well by
\begin{eqnarray}\label{eq:dispersionApx_1}
E_{\pm}&\approx&\sqrt{\xi^2_{\pm}+\left(\Delta_0\pm\frac{\lambda}{|\vec{B}_\bk|}\Delta_2\right)^2},
\end{eqnarray}
where the band dispersion has the form
\begin{eqnarray}\label{eq:band dispersion_1}
\xi_{\pm}&=&A_{\bk}\pm |\vec{B}_\bk|=\mu-\frac{\bk^2}{2m}\pm\sqrt{R_\theta \frac{\bk^4}{4m^2}+\lambda^2}.
\end{eqnarray}
As mentioned in the main text, we parameterize $R_\theta$  as
\begin{eqnarray}\label{eq:Rtheta_1}
R_\theta = p_0 \left(\frac{1}{2} + p_1 + \left(\frac{1}{2} - p_1\right) \cos4\theta\right),
\end{eqnarray}
where, without loss of generality, $0<p_0<1$ and $0<p_1<\frac{1}{2}$.

The top (bottom) sign in (\ref{eq:band dispersion_1}) corresponds to the outer (inner) Fermi surface.
Eq.(\ref{eq:dispersionApx_1}) corresponds to the dashed line in the right panel of Fig.1 of main text, which as we see, captures well the low energy avoided level crossing caused by pairing. It misses the level crossing at high energy,
but at weak coupling this is unimportant.
The form of the Eq.(\ref{eq:band BdG_1}) makes it clear that for $\Delta_0\rightarrow 0$, the pairing is of $s_{\pm}$ nature. Moreover, it is anisotropic, because the zeros of $\xi_{\pm}$ do not coincide with the minima of $|{\vec B}_\bk|$. In addition, the gap is smaller on the outer Fermi surface, because the factor $|{\vec B}_\bk|$ is larger there.

At weak coupling, when $T_c \ll \mu$ even if $\lambda/\mu = O(1)$, $|\Delta_0| \ll |\frac{\lambda}{|\vec{B}_\bk|}\Delta_2|$.  In this situation,
one can approximate the gaps $\Delta_{\pm}$ by $\pm \frac{\lambda}{|\vec{B}_\bk|}\Delta_2$.
The values of $\Delta_+$ and $\Delta_-$ are universally expressed via $T_c$.  The expressions are particularly simple for
 $p_1=1/2$, when $R_\theta =p_0$ reduces to a constant. In this case we have
 \begin{eqnarray}\label{eq:new1}
 && \left| \frac{\Delta_-}{\Delta_+}\right| = \frac{\sqrt{p_0 + (1-p_0) \frac{\lambda^2}{\mu^2}} + p_0}{\sqrt{p_0 + (1-p_0) \frac{\lambda^2}{\mu^2}} - p_0},  \\
 &&\left|\Delta_+\right|^{\frac{|\Delta_-|}{|\Delta_-|+|\Delta_+|}}  \left|\Delta_-\right|^{\frac{|\Delta_+|}{|\Delta_-|+|\Delta_+|}} = \Delta_{BCS},
 \end{eqnarray}
 where $\Delta_{BCS} = 1.76 T_c$.

\subsection{The gap ratio}

We now keep $\Delta_0$ in the gap equation and compute the ratio $K(T) = \Delta_0 (T)/\Delta_2 (T)$.
At weak coupling, when $\Delta_2 \ll \mu$,  the calculation of the gap ratio at $T=T_c$ yields
 $K(T_c) \ll 1$.  In this limit,  we found analytically
\begin{eqnarray}
K(T=0)
\approx  K(T_c) (1 + \mathcal{A}),
\end{eqnarray}
where
\begin{equation}
\mathcal{A}=  \left(1+\frac{\delta\chi_{02}}{\chi_{02}\left(T_c\right)}+\ldots \right).
\label{eq:new2}
\end{equation}
In (\ref{eq:new2})
\begin{equation}
\chi_{02}(T_c)=\frac{m}{2\pi}\frac{\lambda}{\mu}\int_0^{2\pi}\frac{d\theta}{2\pi}
\frac{\tanh^{-1}\sqrt{R_\theta+(1-R_\theta)\frac{\lambda^2}{\mu^2}}}
{\sqrt{R_\theta+(1-R_\theta)\frac{\lambda^2}{\mu^2}}},
\label{f_4a}
\end{equation}
and
\begin{eqnarray}
\delta \chi_{02}&=&\frac{m}{2\pi}\frac{\lambda}{\mu}\int_0^{2\pi}\frac{d\theta}{2\pi}
\frac{\ln\frac{1+\mathcal{F}(\theta,\frac{\lambda}{\mu})}{1-\mathcal{F}(\theta,\frac{\lambda}{\mu})}}
{\sqrt{R_\theta+(1-R_\theta)\frac{\lambda^2}{\mu^2}}},
\end{eqnarray}
where $\mathcal{F}(\theta,\frac{\lambda}{\mu})=R_\theta/\sqrt{R_\theta+(1-R_\theta)\frac{\lambda^2}{\mu^2}}<1$.
Because $\delta\chi_{02}$ and $\chi_{02}\left(T_c\right)$  have the same sign (the same as the sign of $\lambda$), their ratio is positive, hence $\mathcal{A} >0$, i.e., the magnitude of
$\Delta_0/\Delta_2$ is larger at $T=0$ than at $T_c$.

For circular pockets, when $p_1=1/2$ and $R_\theta = p_0$, and for $\lambda \ll \mu$, we obtained a very simple result: $\mathcal{A} =2$, i.e., the ratio $\Delta_0/\Delta_2$ at $T=0$ is three times larger than at $T_c$.


\begin{thebibliography}{10}
\bibitem{comm_n} More recent theoretical studies ~\cite{si,we_last} pointed out that a full gap at the FS wavevectors does not necessarily rule out $d-$wave because the gap minimum does not reside along the normal state FS.
    Yet,   for KFe$_{2}$As$_{2}$ the measured
  angle variation of the gap on the inner pocket is still much weaker than the one expected for a d-wave gap~\cite{we_last}.
\end{thebibliography}
\end{document}